# Understanding transient uncoupling induced synchronization through modified dynamic coupling

Anupam Ghosh,[1, a)] Prakhar Godara,[1, b)] and Sagar Chakraborty[1, c)]
*Department of Physics, Indian Institute of Technology Kanpur, Uttar Pradesh 208016, India*

An important aspect of the recently introduced transient uncoupling scheme is that it induces synchronization for large values of coupling strength at which the coupled chaotic systems resist synchronization when continuously coupled. However, why this is so is an open problem. To answer this question, we recall the conventional wisdom that the eigenvalues of the Jacobian of the transverse dynamics measure whether a trajectory at a phase point is locally contracting or diverging with respect to another nearby trajectory. Subsequently, we go on to highlight a lesser appreciated fact that even when, under the corresponding linearised flow, the nearby trajectory asymptotically diverges away, its distance from the reference trajectory may still be contracting for some intermediate period. We term this phenomenon transient decay in line with the phenomenon of the transient growth. Using these facts, we show that an optimal coupling region, i.e., a region of the phase space where coupling is on, should ideally be such that at any of the constituent phase point either the maximum of the real parts of the eigenvalues is negative or the magnitude of the positive maximum is lesser than that of the negative minimum. We also invent and employ modified dynamics coupling scheme—a significant improvement over the well-known dynamic coupling scheme—as a decisive tool to justify our results.

**Owing to the sensitive dependence on initial conditions, identical synchronization of two coupled chaotic systems, the drive and the driven, appears very counterintuitive. It is additionally intriguing to note that occasional switching off of the coupling between the drive and the driven systems may lead to synchronization when continuous coupling fails. Here we find the mechanism behind the success of one such occasional uncoupling scheme, the transient uncoupling scheme, in which the coupling is active only when the phase trajectory visits a fixed fraction of the corresponding attractor. The mechanism essentially showcases that the success of such a scheme hinges on the fact whether it can eliminate the locally diverging dynamics in the phase space of the error variables—the differences between the variables of the drive and the driven systems.**

## I. INTRODUCTION

From the metabolic processes in our cells[1] to the extended ecological systems[2], synchronization is one of the most common phenomenon in nature. Although synchronization in chaotic systems had been discussed by others, e.g., Fujisaka and Yamada[3], and Afraimovich et al.[4], it was the work of Pecora and Carroll[5,6] on synchronization of chaotic systems that caught the wide-spread attention of the researchers. Further studies resulted in the realization that there are many different types of synchronization of chaotic systems, viz., identical synchronization[5], generalized synchronization[7], phase synchronization[8], lag synchronization[9] etc.

It is obvious that coupling is requisite to synchronize oscillators, but further developments in the field of chaotic synchronization have revealed that sometimes uncoupling the systems helps make the synchronization more robust. In such schemes, the response/driven subsystem does not get the drive signal continuously with time[10]; the response variables are updated occasionally while being directly coupled with the drive variables and in between two such consecutive updations, the response variables evolve independently uncoupled from the drive subsystem. These occasional coupling synchronization schemes require lesser transfer of information between the interacting subsystems compared to when they are continuously coupled. Also, they have recently been known to be used to synchronize unsynchronizable networks[11]. Amritkar and Gupte[12] were among the first ones to discover the idea of occasional driving synchronization of chaotic systems. Other such schemes are on-off coupling scheme[13], dynamic coupling scheme[14], transient uncoupling scheme[15], and sporadic driving scheme[16,17]. While, in this paper, we discuss the first three schemes as and when need arises, *the main question we address is: how diffusively coupled chaotic systems synchronize when transiently uncoupled*? Although a heuristic explanation has been supplied by Tandon and coworkers[18] in the context of generalized synchronization, it is not satisfactory and is not strictly applicable for very high coupling strengths as shown in this paper. In passing, we remark that while we are exclusively concerned with deterministic schemes, successful occasional coupling schemes involving stochastic switching of coupling is also well-known in the literature[19–22].

Before we present the results of our investigation, let us start with closely scrutinizing the transient uncoupling

[a)] Electronic mail: anupamgh@iitk.ac.in
[b)] Electronic mail: prakhg@iitk.ac.in
[c)] Electronic mail: sagarc@iitk.ac.in

scheme in the immediately following section.

## II. SCRUTINY OF TRANSIENT UNCOUPLING SCHEME

The general form of two identical chaotic oscillators under unidirectionally diffusive transient uncoupling scheme is

$$\frac{d\mathbf{x}_1}{dt} = \mathbf{F}(\mathbf{x}_1), \tag{1a}$$

$$\frac{d\mathbf{x}_2}{dt} = \mathbf{F}(\mathbf{x}_2) + \alpha \chi_\mathbb{A}(\mathbf{x}_2) \mathsf{C} \cdot (\mathbf{x}_1 - \mathbf{x}_2), \tag{1b}$$

where $\mathbf{x}_1(t)$ and $\mathbf{x}_2(t)$ are the states of the drive and the response subsystems respectively in the corresponding $d$-dimensional phase spaces, $\mathbb{R}^d$. The matrix $\mathsf{C}$ is the coupling matrix (a square matrix) and the scalar $\alpha$ is the coupling strength that measures the strength of the coupling. $\chi_\mathbb{A}(\mathbf{x}_2)$ imparts the transient uncoupling and is defined as:

$$\chi_\mathbb{A}(\mathbf{x}_2) := \begin{cases} 1 \text{ for } \mathbf{x}_2 \in \mathbb{A}, \\ 0 \text{ for } \mathbf{x}_2 \notin \mathbb{A}. \end{cases} \tag{2}$$

In other words, $\mathbb{A}$, the coupling region, is a subset of the driven subsystem's phase space $\mathbb{R}^d$ and the coupling is active only when the phase space trajectory is within $\mathbb{A}$.

At stable synchronized state $\mathbf{x}_1(t) = \mathbf{x}_2(t) = \mathbf{x}_s(t)$ (say), the dynamics of the transverse state $\mathbf{x}_\perp(t) =: \mathbf{x}_1(t) - \mathbf{x}_2(t)$ should converge to $\mathbf{x}_\perp(t) \to 0$. The necessary condition ensuring this is that the maximum transverse Lyapunov exponent $\lambda_{\max}^\perp =: \lim_{t\to\infty}(1/t)\ln[||\mathbf{x}_\perp(t)||/||\mathbf{x}_\perp(0)||]$ (where $||\cdots||$ is the standard Euclidean norm) should be negative for the linearized transverse dynamical system:

$$\dot{\mathbf{x}}_\perp \approx \tilde{\mathsf{J}}(\mathbf{x}_s, \alpha; \mathbb{A})\mathbf{x}_\perp =: [\partial_\mathbf{x}\mathbf{F}(\mathbf{x}_s) - \alpha\chi_\mathbb{A}(\mathbf{x}_s)\mathsf{C}]\mathbf{x}_\perp. \tag{3}$$

We call $\tilde{\mathsf{J}}(\mathbf{x}_s, \alpha; \mathbb{A})$ the effective Jacobian of the coupled system for obvious reasons.

Although it is not necessary at all to work with the Rössler oscillator[23], in this paper we work with it as a convenient prototype system. There are many other systems, e.g., Lorenz system[24] and Chen system[25], on which the transient uncoupling scheme can be successfully employed[11,15,18]. The Rössler oscillator is mathematically represented explicitly by the following three dimensional autonomous dynamical system:

$$\dot{x}_1 = -x_2 - x_3, \tag{4a}$$
$$\dot{x}_2 = x_1 + ax_2, \tag{4b}$$
$$\dot{x}_3 = b + x_3(x_1 - c). \tag{4c}$$

We take parameters $a = b = 0.2$ and $c = 5.7$ so that the oscillator shows chaotic behaviour. Using x-coupling (i.e., when $\mathsf{C}_{11} = 1$ is the only non-vanishing element of the matrix) and a coupling region chosen as

$$\mathbb{A} = \{\mathbf{x}_2 \in \mathbb{R}^3 : ||(\mathbf{x}_2)_1 - (\mathbf{x}_2^*)_1|| \leq \Delta\}, \tag{5}$$

with the parameter values $\Delta = 4.16$, $(\mathbf{x}_2^*)_1 = 1.20$[15], allows the diffusively coupled Rössler systems to synchronize even for large $\alpha$ values ($\alpha \gtrsim 4.3$) unlike the continuous coupling scheme where the synchronization happens only in the range $0.2 \lesssim \alpha \lesssim 4.3$.

Although the choice of $\mathbb{A}$ is an exercise in trial and error, a later paper[18] provides following reason justifying the choice of $\mathbb{A}$: consider the eigenvalues—$\lambda_1$, $\lambda_2$, and $\lambda_3$—of the effective Jacobian $\tilde{\mathsf{J}}(\mathbf{x}_s, \alpha; \mathbb{R}^3)$ and assume, without any loss of generality, $\text{Re}(\lambda_1) \geq \text{Re}(\lambda_2) \geq \text{Re}(\lambda_3)$. For later convenience, we define $\Lambda_{\max} =: \text{Re}(\lambda_1)$ and $\Lambda_{\min} =: \text{Re}(\lambda_3)$. The simple idea is to choose $\mathbb{A}$ in such a fashion that it contains those phase points, at which $\Lambda_{\max} \leq 0$, in the attractor of the driven subsystem.

In Fig. 1, we exhibit the projected phase space attractor of the driven subsystem. In the figure, the coupling region is the region sandwiched between the vertical dashed lines and we see how the number of phase points with $\Lambda_{\max} \leq 0$ inside the region first increases and then decreases with the coupling strength.

We straightaway note that this choice of $\mathbb{A}$, as suggested above, should depend on value of $\alpha$ being tried in corresponding continuous coupling scheme. However, the transient uncoupling scheme advocates for a fixed coupling region irrespective of the value of $\alpha$. While this may work, it is certainly not the most optimal solution. As an example, we note that the specific coupling region defined in Eq. (5), although is appropriate for bringing the Rössler systems to synchronize for higher values of coupling strength, can not impart synchronization for the lower range of $\alpha$ values: $0 < \alpha \lesssim 0.35$.

In passing, we may remark here that if one calculates the error-bars on $\lambda_{\max}^\perp$ (that we have done using 700 different initial conditions), then the upper bound 0.35 doesn't appear very strict; for quite a few initial conditions, $\lambda_{\max}^\perp$ can be zero even at $\alpha = 0.22$. This value is almost equal to the similar upper bound on the coupling strength, i.e., $\alpha \approx 0.2$ ($\approx 0.21$ when error-bars are calculated), for the continuous coupling case. The error-bar mentioned above very generously refers to the spread of only one standard deviation, $\sqrt{\langle[\lambda_{\max}^\perp - \langle\lambda_{\max}^\perp\rangle]^2\rangle}$, on each side about the mean, $\langle\lambda_{\max}^\perp\rangle$. The average $\langle\cdots\rangle$ has been defined over the initial conditions. Similar conclusions can be made for the other threshold of synchronization, viz., $\alpha = 4.3$, for the case of continuous coupling. Thus, the thresholds of synchronization are, in practice, almost always uncertain and all our discussions in this paper are for the coupling strengths sufficiently away from these uncertain bands.

Coming back to the main discussion, a far more serious issue is that beyond the synchronization threshold, say, $\alpha \gtrsim 4.5$ there are almost no phase points with $\Lambda_{\max} \leq 0$, but still we know that[15] the transient uncoupling is capable of imparting synchronization. This is directly at odds with the aforementioned prescription of choosing the coupling region. So what causes the transient uncoupling to be effective for larger values of $\alpha$ is not clear a

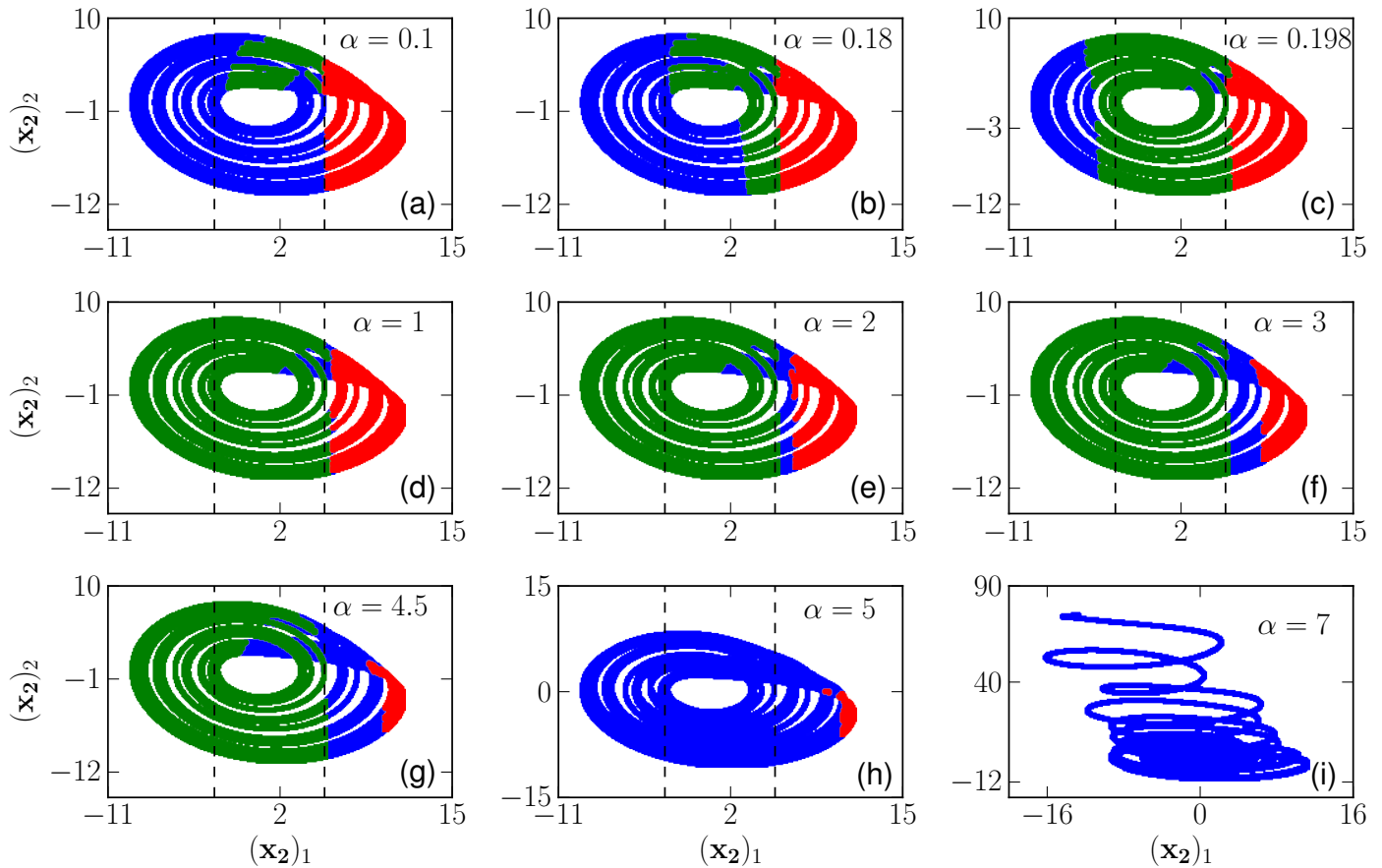

FIG. 1. *(Color online)* Plots of the attractors of the driven subsystem for different values of the coupling strength ($\alpha$) when $x$-coupled Rössler oscllators are continuously coupled. While the green colour corresponds to the points at which $\Lambda_{\max} < 0$, the other two colours, viz., blue and red, collectively correspond to the points at which $\Lambda_{\max} > 0$. It is of convenience for the discussion in Sec. IV to note that while the blue refers to the points with the additional feature that $|\Lambda_{\max}| < |\Lambda_{\min}|$, the red does so for the points with $|\Lambda_{\max}| > |\Lambda_{\min}|$. Also, note how for larger strengths of coupling the response subsystem diverges. In this paper, the coupling region for employing the transient uncoupling scheme is so chosen that $(\mathbf{x}_2^*)_1 = 1.20$ and $\Delta = 4.16$[15]. This region is shown bounded between the two vertical black dashed lines. Subfigures (a), (b), (g), and (i), with $R_1$ values 6.80, 21.97, 0.0, and 0.0 respectively, correspond to unsynchronized states. In contrast, subfigures (c)-(g) correspond to synchronized states for which the values of $R_1$ are relatively much higher and they specifically are 60.52, 80.78, 80.38, 79.73, and 76.99 respectively.

*priori*.

The two immediately preceding paragraphs make it crystal clear that *why the transient uncoupling works as it does is an ill-understood problem* and this is the motivation behind the investigation presented herein. For further discussion we define a parameter,

$$R_1 = 100 \times \frac{N_-}{N_- + N_+}, \quad (6)$$

where, $N_-$ is the total number of phase points with negative $\Lambda_{\max}$ and $N_+$ is the total number of phase points with positive $\Lambda_{\max}$. Although, there are issues as discussed earlier, large $R_1$ values helps synchronization and the smaller values of $R_1$ would mean desynchroinzation. This is at one with Fig. 1 which pictorially shows how a relative increase in the values of $R_1$ (see the figure's caption also) in the range $0.2 \lesssim \alpha \lesssim 4.5$ goes hand in hand with the existence of the stable synchronized states in that range for continuous coupling. In fact, at higher $\alpha$, $R_1 = 0$ as the response subsystem becomes unstable and diverges. This is not very surprising as it has been analytically shown[10] for very large coupling strengths.

Interestingly, there is a third caveat. The sign of $\Lambda_{\max}$ found using $\mathsf{J}(\mathbf{x_s}, \alpha; \mathbb{R}^3)$ actually tells us about the continuous coupling induced synchronization and we are supposed to pick the coupling region for the transient uncoupling scheme by looking at this very data. Idea is that the regions with a large fraction of points with $\Lambda_{\max} < 0$ should be used as the coupling region while employing the transient uncoupling. However, the fact is that when the transient uncoupling is in action, the set of points where the trajectories are locally contracting and the set of points where the trajectories are locally expanding may not be identical to the respective corresponding sets when continuous coupling is in action. With little thought, one can appreciate that although for large $\alpha$ (say, $\gtrsim 5$), $R_1 = 0$ for continuous coupling, it cannot be so when the transient uncoupling is in action, simply because otherwise the transient uncoupling cannot induce synchronization contrary to what is factually true. It is however interesting to know that the corresponding non-zero value of $R_1$ is very small.

## III. MODIFIED DYNAMIC COUPLING SCHEME

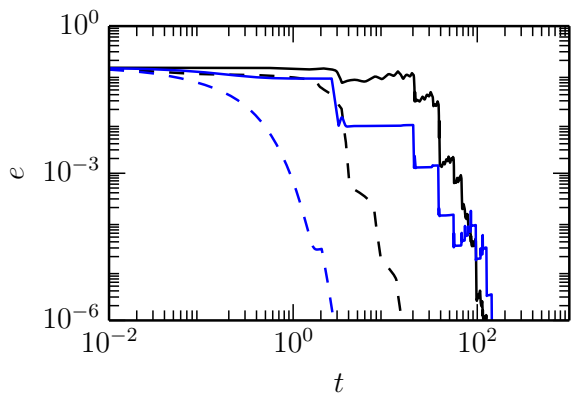

FIG. 2. *(Color online)* Success of the modified dynamic coupling scheme. The variation of the synchronization error ($e = ||\mathbf{x}_1 - \mathbf{x}_2||$) with time ($t$) is showing that the $x$-coupled Rössler oscillators are synchronized using both the dynamic coupling scheme (dashed lines) and the modified dynamic coupling scheme (solid lines). Both the schemes are plotted for two different $\alpha$ values: the black and the blue colours correspond to $\alpha = 0.1$ and $\alpha = 5.0$ respectively.

We reconsider Eqs. (1a)-(1b) that with $\mathbb{A} = \mathbb{R}^3$ means continuous coupling. As detailed in the last section, when e.g., $\alpha = 0.1$ or $\alpha = 5.0$, two such coupled Rössler systems are not in synchrony. We now show that the idea behind the dynamics coupling scheme[14], adapted appropriately for the problem in hand, can induce synchronization. Subsequently, we go on to modify the scheme—terming the newer scheme as the modified dynamics coupling scheme —to conclusively establish that $\Lambda_{\min}$ must also be considered along with $\Lambda_{\max}$ in order to figure out when the synchronization be effected.

The idea behind the dynamic coupling scheme in the context of our system is best explained by writing the following expressions:

$$\frac{d\mathbf{x}_1}{dt} = \mathbf{F}(\mathbf{x}_1), \quad (7a)$$

$$\frac{d\mathbf{x}_2}{dt} = \mathbf{F}(\mathbf{x}_2) + \alpha\mathsf{C} \cdot (\mathbf{x}_1 - \mathbf{x}_2) + \mathsf{C}_{\mathrm{dcs}}(\mathbf{x}_1 - \mathbf{x}_2), \quad (7b)$$

where $\alpha = 0.1$ or 5 and $\mathsf{C} = \mathrm{diag}(1, 0, 0)$. Note that $\mathsf{C}_{\mathrm{dcs}}$ is not a constant matrix like $\mathsf{C}$ and is a function of the phase space variables. When the synchronization error $\mathbf{x}_\perp$ evolves in the linear regime as $\dot{\mathbf{x}}_\perp = \tilde{\mathsf{J}}(\mathbf{x_s}, \alpha; \mathbb{R}^3)\mathbf{x}_\perp$, the possible local divergence of the trajectories in the phase space—as characterized by positive $\Lambda_{\max}$—can be tackled by the additional $\mathsf{C}_{\mathrm{dcs}}(\mathbf{x}_1 - \mathbf{x}_2)$ term. $\mathsf{C}_{\mathrm{dcs}}$ is chosen in such a manner that the synchronization error, $e := ||\mathbf{x_1} - \mathbf{x_2}||$, always decreases with time. The detailed theory of how to choose $\mathsf{C}_{\mathrm{dcs}}$ may be gathered from the original paper[14] on the scheme. However, for the sake of completeness, we now briefly discuss the dynamic coupling scheme as adapted for our problem: If we start with initial condition $\mathbf{x}_\perp(0)$, after an infinitesimal time interval $\Delta t$, we have the approximation: $\mathbf{x}_\perp(\Delta t) = \mathsf{L}(\Delta t) \cdot \mathbf{x}_\perp(0)$, where $\mathsf{L}(\Delta t) := (\mathsf{I} + \Delta t \tilde{\mathsf{J}})$ and $\mathsf{I}$ is the identity matrix. Thus,

$$\mathsf{L}^{tr}(\Delta t) \cdot \mathsf{L}(\Delta t) \approx \mathsf{I} + \Delta t(\tilde{\mathsf{J}}^{tr} + \tilde{\mathsf{J}}) = \mathsf{I} + \Delta t \mathsf{M}, \quad (8)$$

after neglecting the higher order terms of $\Delta t$. Here, superscript '$tr$' stands for the transpose and $\mathsf{M} := (\tilde{\mathsf{J}}^{tr} + \tilde{\mathsf{J}})$. On using the singular value decomposition, we can write $\mathsf{L}$ as $\mathsf{L} = U \cdot W \cdot V^{tr}$, where $\mathsf{U}$ and $\mathsf{V}$ are orthogonal matrices and $\mathsf{W}$ is a diagonal matrix with elements $w_i$, here $i = 1, 2, 3$. Thus, it readily follows that $\mathsf{M} = \mathsf{V} \cdot \mathsf{D} \cdot \mathsf{V}^{tr}$, where $\mathsf{D} = (\mathsf{W}^2 - \mathsf{I})/\Delta t$ is a diagonal matrix with diagonal elements $d_i$ and, of course, $w_i = \sqrt{1 + \Delta t d_i}$. Obviously, if $w_i < 1$ (or equivalently $d_i < 0$) then the local contraction is possible. With this in mind, the coupling matrix $\mathsf{C}_{\mathrm{dcs}}$, by construction, is chosen to be a symmetric matrix such that $\mathsf{C}_{\mathrm{dcs}} + (\mathsf{C}_{\mathrm{dcs}})^{tr} = 2\mathsf{V} \cdot \mathrm{diag}(c_1, c_2, c_3) \cdot \mathsf{V}^{tr}$ with $c_i > d_i/2$ making the matrix $\tilde{\mathsf{J}} - \mathsf{C}_{\mathrm{dcs}} + (\tilde{\mathsf{J}} - \mathsf{C}_{\mathrm{dcs}})^{tr}$ to possess only negative eigenvalues.

Before proceeding further, we would like to stress that in the original paper[14] on the dynamical coupling, the term $\alpha\mathsf{C} \cdot (\mathbf{x}_1 - \mathbf{x}_2)$ is not considered as the aim of that paper is different; in fact, the original aim behind the dynamic coupling was to replace the continuous coupling by a dynamically changing coupling that always induces synchronization. Here, in contrast, we are dynamically changing an *additional* coupling term, $\mathsf{C}_{\mathrm{dcs}}(\mathbf{x}_1 - \mathbf{x}_2)$, in order to impart synchrony when mere continuous coupling with constant strength fails.

We employ the dynamic coupling, i.e., turn on $\mathsf{C}_{\mathrm{dcs}}$, whenever $\Lambda_{\max} > 0$ and not otherwise. As seen in Fig. 2, the synchronization error goes to vanishingly small values with time. It should be noted that at $\alpha = 0.1$, the dynamic coupling is employed for about 93% of the points (the non-green region in Fig. 1a) on the attractor while at $\alpha = 5$, when $R_1 = 0$, naturally the dynamic coupling is employed for 100% of the phase space attractor. Of course, these percentages are calculated after removing same lengths of transients in each case.

We foresee that the above scheme can be modified so as to employ the dynamic coupling for far less number of points in the phase space by taking into account the following simple insight: Consider, for convenience of argument, a two dimensional phase space $\mathbb{R}^2$ in which a trajectory ($\mathbf{x}_t$) is being traced out by a two dimensional autonomous dynamical flow. Assume there exists a saddle (fixed) point ($\mathbf{x}^\star$), meaning that the Jacobian of the linearized equation has one negative eigenvalue ($\sigma_-$) and one positive eigenvalue ($\sigma_+$). Any initial condition ($\mathbf{x}_0$) in the vicinity of the saddle, but not on the stable or the unstable manifold, tries to approach the saddle in the direction parallel to the stable eigendirection and simultaneously go away from the saddle in the direction parallel to the unstable eigendirection. If $|\sigma_+| < |\sigma_-|$, then for some time the distance between the saddle and the trajectory, $||\mathbf{x}^\star - \mathbf{x}_t||$, can be less than the distance between the saddle and the initial condition, $||\mathbf{x}^\star - \mathbf{x}_0||$. Of course, eventually $||\mathbf{x}^\star - \mathbf{x}_t|| \to \infty$. Intriguingly, this decrease in $||\mathbf{x}^\star - \mathbf{x}_t||$ for a small span of time may be contrasted with the phenomenon of transient growth[26] to realize that both the phenomena are analogous; in the latter, there is growth of separation between a stable node and a nearby trajectory before the trajectory reaches the fixed point asymptotically. Thus, in analogy with the transient growth, the former phenomenon may be termed as transient decay.

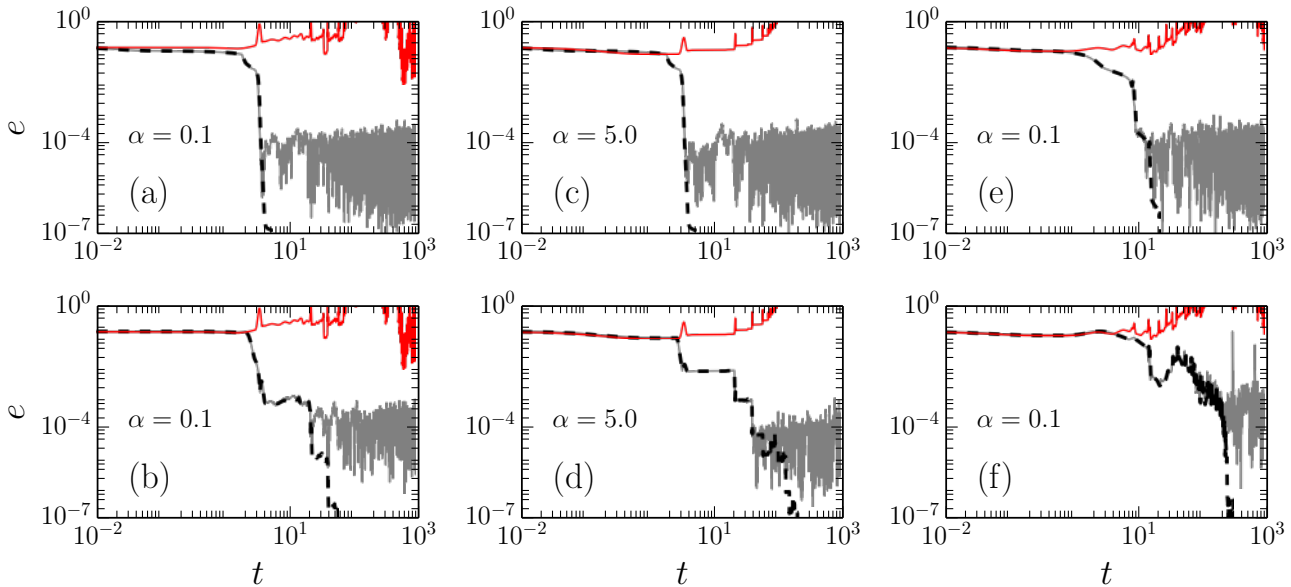

FIG. 3. *(Color online)* Robustness of the dynamic coupling scheme and the modified dynamic coupling scheme against noise. Subplots (a), (b), (c) and (d), correspond to $x$-coupled Rössler oscillators at two different coupling strengths $\alpha = 0.1$ and $\alpha = 5.0$ as labeled therein, whereas subplots (e) and (f) correspond to $x$-coupled hyperchaotic Rössler oscillators at the coupling strength $\alpha = 0.1$. Upper subplots (a), (c), and (e) show how the dynamic coupling imparts synchronization in the otherwise non-synchronizable cases; red solid curves, black dashed curves, and grey solid curve respectively correspond to the continuously coupled case, the dynamically coupled case, and the dynamically coupled case where noised has been added. Similarly, lower subplots (b), (d), and (f) depict how the modified dynamic coupling induces synchronization in the otherwise non-synchronizable cases; red solid curves, black dashed curves, and grey solid curve respectively correspond to the continuously coupled case, the modified dynamically coupled case, and the modified dynamically coupled case where noised has been added.

Equipped with this insight, we propose what we call the modified dynamic coupling scheme, where we predict that rather than keeping dynamic coupling $\mathsf{C}_{\text{dcs}}$ turn on whenever $\Lambda_{\max} > 0$, we keep it on only when $|\Lambda_{\max}| >$

$|\Lambda_{\min}|$ at a phase point. To emphasize again, may be redundantly, we mention that the dynamic coupling is turned off at a point if $|\Lambda_{\max}| < |\Lambda_{\min}|$ ($\Lambda_{\min} < 0$), even though $\Lambda_{\max} > 0$. In other words, we are claiming that the transient decay contributes as locally contracting dynamics even though the positivity of $\Lambda_{\max}$, when taken literally, suggests otherwise. We test our claim on our system, i.e., diffusively coupled Rössler oscillator. As depicted in Fig. 2, we find that the modified dynamic coupling is capable of inducing synchronization. More surprising is the fact that the new scheme is only required to be active on $\sim 17\%$ and $\sim 2\%$ of the phase points on the attractor when $\alpha = 0.1$ and $\alpha = 5.0$ respectively. Again, these numbers are evaluated only after getting rid of equal amount of enough transients in both the cases. It is obvious that these two sets of phase points correspond the red regions shown in Fig. 1(a) and Fig. 1(h).

Both the dynamic coupling scheme and the modified dynamic coupling scheme are fairly robust to the noise. To show this, we first introduce a noise term $D\xi(t)$, where $D$ is the noise amplitude and $\xi(t)$ is chosen from a Gaussian random distribution with zero mean and unit variance. The noise is assumed to be uncorrelated, i.e., $\langle \xi(t)\xi(t') \rangle = \delta(t - t')$. $D$ is chosen to be $10^{-4}$ units. We add this noise in $x$-component of the response oscillator of the coupled Rössler oscillators. Fig. 3(a), and (c) exhibit the robustness of the dynamic coupling scheme when $\alpha = 0.1$ and $\alpha = 5.0$ respectively, whereas the other two plots Fig. 3(b), and (d) depict similar robustness of the modified dynamic coupling scheme. We note that the error variable for the synchronized state fluctuates within the amplitude of the noise strength as expected. This robustness against noise is fairly general. In order to illustrate it, two types of the dynamic couplings have been employed on the hyperchaotic Rössler oscillator[27], which is a four dimensional chaotic oscillator. To this end, we take unidirectionally $x$-coupled oscillators and set $\mathbf{F}(\mathbf{x}) = [-y - z, x + ay + w, b + xz, -cz + dw]$, $a = 0.25$, $b = 3$, $c = 0.5$, $d = 0.05$, and the coupling strength $\alpha = 0.1$. Note the appearance of the fourth coordinate, $w$, because this system is four dimensional. Here again, as exhibited by Fig. 3(e) and (f), the dynamic coupling and the modified dynamic coupling induce synchronization in the otherwise desynchronized case (i.e., when continuously coupled); Robustness against noise is also apparent in these plots.

We thus conclude that we have improved significantly over the dynamic coupling scheme. However, this is not the main aim of this section. *In the context of this paper, our useful achievement through the invention of this new scheme is that the magnitude of $\Lambda_{\min}$ is crucial in effecting a stable synchronized dynamics in the diffusively coupled systems.*

## IV. EIGENVALUE WITH SMALLEST REAL PART IS IMPORTANT

Having illustrated that the eigenvalues with smallest real part is important in driving oscillators in synchrony, we now need to appreciate that both the dynamic coupling and the transient uncoupling are partly similar in approach: In the former scheme, one *dynamically* looks for the phase points where the dynamics is locally contracting and leaves the system unmodulated; whereas in the later scheme, one looks to find all the phase points where the maximum of the real parts of the corresponding eigenvalues have negative real part and subsequently chooses a *fixed* coupling region right before employing the scheme such that mostly these points are in the region. Of course, as detailed in the immediately preceding section, in situations where $R_1 = 0$ (e.g., as happens at $\alpha = 5.0$ for diffusively $x$-coupled Rössler oscillators), the choice of coupling region should be motivated by the fact such a region should facilitate transient decay.

However, there is also a drastic difference between the two schemes: While coupling dynamically, on finding a phase point with locally diverging dynamics, one immediately modulates the eigenvalues of the (symmetrized) Jacobian to become negative by introducing additional state variable dependent coupling; during the transient uncoupling, at such a point synchronization is tried by choosing a coupling region that the coupling is simply turned off leading to the completely independent evolu-tions of the drive and the driven subsystems. Thus, the dynamic coupling always succeeds in synchronizing the systems but the transient uncoupling may not. The success of the transient uncoupling hinges on the hope that at such points the eigenvalues are not less favourable to nearby contracting trajectories when the systems are uncoupled than when they are coupled.

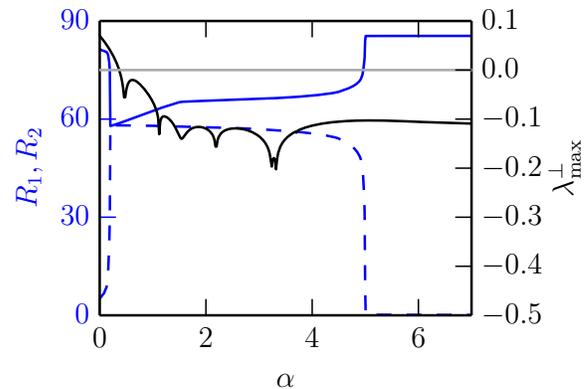

FIG. 4. *(Color online)* Larger $R_2$ compensates small $R_1$ in order to synchronize. The parameters $R_1$ (blue dashed), $R_2$ (blue solid), and the maximum conditional Lyapunov exponent, $\lambda_{\max}^{\perp}$, (black solid) are plotted with coupling strength, $\alpha$, for the $x$-coupled Rössler oscillators employing the transient uncoupling. The gray horizontal line, $\lambda_{\max}^{\perp} = 0$, serves as an aid to eye. The coupling region is so chosen that $(\mathbf{x}_2^*)_1 = 1.20$ and $\Delta = 4.16$[15].

In view of the above, and once we note that the coupling region (see Fig. 1) encompasses exclusively those points at which $|\Lambda_{\max}| < |\Lambda_{\min}|$, it is not surprising that the transient uncoupling brings about synchronization at $\alpha = 5.0$. Note that had we used the modified dynamic coupling in place of the transient uncoupling, the modified dynamic coupling would have left the coupled system unmodified in the coupling region. Once again we are left to appreciate how fortunate we are with this particular choice of coupling region, and how this example validates the partial equivalence between the modified dynamic coupling and the transient uncoupling.

Additionally, it is worth recalling that for higher coupling strengths at which the transient uncoupling synchronizes the systems in hand (refer to Fig. 4), even though $R_1$ (calculated using $\tilde{\mathbb{J}}(\mathbf{x}_s, \alpha; \mathbb{A})$) becomes very small. In this context, using the insight we have gained in this section, we define another parameter:

$$R_2 = 100 \times \frac{\text{Fraction of } N_+ \text{ where } |\Lambda_{\max}| < |\Lambda_{\min}|}{N_+}. \quad (9)$$

It is easy to note that larger the value of $R_2$ is, the more favourable are the conditions for synchronization: $R_2$ increases when the fraction of phase points, where $\Lambda_{\max} > 0$ and $|\Lambda_{\max}| < |\Lambda_{\min}|$, increases with respect to the similar fraction having $|\Lambda_{\max}| > |\Lambda_{\min}|$. Fig. 4 shows that $R_2$ (also calculated using $\tilde{\mathsf{J}}(\mathbf{x}_s, \alpha; \mathbb{A})$) being relatively more makes synchronization possible, although $R_1$ is almost vanishingly small.

## V. COMPARING TRANSIENT UNCOUPLING SCHEME AND ON-OFF COUPLING SCHEME

Before we conclude, we compare the transient uncoupling scheme with another closely related occasional coupling scheme called the on-off coupling scheme. In the on-off coupling scheme[13], the coupling is alternatively active (on) and inactive (off) for some fixed specific fractions of time. Mathematically, if $T$ is the on-off period and $\theta$ is the on-off rate then for the duration $nT < t < (n+\theta)T$ (where $n = 0, 1, 2, \cdots$ and $\theta \in [0,1]$), the coupling is active; for the rest of the time interval $(n+\theta)T < t < (n+1)T$, the coupling is inactive. Thus, the subsystems—the drive and the driven—couple and uncouple in a strictly periodic manner. It is obvious that in contrast to the transient uncoupling scheme (where the coupling term is switched on or off depending on where the trajectory is in the phase space), in the on-off coupling scheme, the coupling is on or off depending on whether the time-coordinate of the corresponding phase point is in $nT < t < (n+\theta)T$ or not. For more explicit comparison with the transient uncoupling scheme, we can write the general form of two coupled oscillators with the on-off coupling, in line with equations (1a) and (1b), as:

$$\frac{d\mathbf{x}_1}{dt} = \mathbf{F}(\mathbf{x_1}), \tag{10a}$$

$$\frac{d\mathbf{x}_2}{dt} = \mathbf{F}(\mathbf{x}_2) + \alpha \tilde{\chi}_{T,\theta}(t) \mathsf{C} \cdot (\mathbf{x}_1 - \mathbf{x}_2), \tag{10b}$$

with

$$\tilde{\chi}_{T,\theta}(t) = \begin{cases} 1 \text{ for } nT < t < (n+\theta)T, \\ 0 \text{ for } (n+\theta)T < t < (n+1)T. \end{cases} \tag{11}$$

While usage of the average inter-peak time-interval of $x$-coordinate time-series as the measure of $T$ has been advocated[28], the choice of the parameter $\theta$ is totally *ad hoc* which is a drawback. Take for example our case of the $x$-coupled Rössler oscillators. Here, the average inter-peak time-interval of the $x$-coordinate is approximately 5.85. Now, using $T = 5.85$ and $\theta = 0.5$ (arbitrarily chosen), synchronized state is not observed at coupling strength $\alpha = 5$. Again, for the $x$-coupled hyperchaotic Rössler oscillators, discussed earlier, the situation is even worse: here, we have checked that using the average inter-peak time-interval $T = 5.18$ doesn't lead to synchronization at $\alpha = 0.1$ for any value of $\theta$. Moreover, the choice of $T$ is also not full-proof as discussed later in this section. While for the on-off coupling one has to specify two unknown parameters ($T$ and $\theta$), for the transient uncoupling one has to specify only the coupling region, $\mathbb{A}$. One may argue that in order to pick optimal $\mathbb{A}$, even before one can employ the method for synchronizing, one needs to solve the system fully to find the eigenvalue spectra of the local Jacobians. However, knowing the solution of the system beforehand is also essential for the on-off coupling as the solution is needed to calculate the average inter-peak time interval and, thus, $T$. In passing, we remark that, in our numerical experiments, the on-off coupling scheme appears to be two-three times faster than the transient uncoupling scheme.

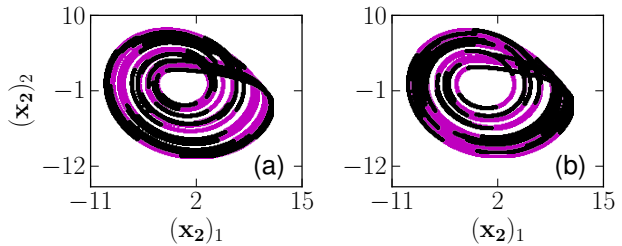

FIG. 5. *(Color online)* Intermingled and non-invariant on-region and off-region in the on-off coupling scheme. This figure exhibits the regions of the driven subsystem's phase space where the coupling is active (magenta) and inactive (black) when the on-off coupling is in action. Subplots (a) and (b) respectively correspond to two different initial conditions, $(\mathbf{x}_1(0), \mathbf{x}_2(0)) = (-9, 0, 0, -9.1, 0.1, 0)$ and $(\mathbf{x}_1(0), \mathbf{x}_2(0)) = (-4, 0.5, 0, -4.1, 0.6, 0)$, taken at time $t = 0$. Equal lengths of transients have been removed before plotting the attractors. Here $T = 3$ and $\theta = 0.5$ (at $\alpha = 5$) so that the on-off coupling leads to the synchronization of the $x$-coupled Rössler oscillators which are otherwise not synchronized when continuously coupled.

It is obvious that even in the case of the on-off coupling, there corresponds a definite non-zero measurable set of points in the phase space where the coupling is on. In this respect, the on-off coupling seems to be nothing but the transient uncoupling, as employed in Eqs. (1a)-(1b), but now choice of $\mathbb{A}$ is not convenient. As we elaborate below, this inconvenience is partly due to the fact that the coupling region may not be connected and partly due to the fact that, by construction of the on-off coupling scheme, this on-region (and the mutually exclusive complementary off-region) is not robust to the changes in the initial conditions, $(\mathbf{x}_1(0), \mathbf{x}_2(0))$, even after getting rid of the transients.

To illustrate this, we perform the on-off coupling scheme on the same $x$-coupled Rössler systems with $\alpha = 5$ (thus, continuously coupled systems are not synchronized) and by *trial-and-error*, find out that $T = 3$ and $\theta = 0.5$ synchronizes the coupled system. However, if we look at the projected two-dimensional phase space portrait of the driven's attractor (see Fig. 5), we note that the set of points at which the coupling is on is haphazardly (at least visually) intermingled with the rest of the trajectory. Further, due to the chaotic (non-periodic) nature of the trajectories, the regions over which the coupling term is active are completely different for any two different sets of initial conditions chosen on the trajectory. It is, thus, clearly impossible to find out a general combination of $T$ and $\theta$ such that the coupling region is given by a simple expression for $\mathbb{A}$ such as the one given by Eq. (5). In other words, even if we know full information about the eigenvalues of $\tilde{\mathsf{J}}(\mathbf{x}_s, \alpha; \mathbb{R}^3)$ and thus, a straightforward connected set $\mathbb{A}$, we cannot *a priori* find out the specific values of $T$ and $\theta$ such that most of the points where coupling is on are elements of the set $\mathbb{A}$. This is not very surprising because a chaotic system is not a periodic system and hence any scheme, like the on-off coupling scheme, that picks points in periodic fashion on the chaotic attractor ends up picking the points irregularly throughout the attractor. Strictly speaking, the heuristic suggestion[28] that $T$ can be considered as the average inter-peak time-interval of $x$ time-series also doesn't work. Moreover, that the on-region and the off-region are not invariant with respect to the changes in initial conditions is crystal clear. In short, evidently, *the on-off coupling scheme is far more an ad hoc scheme than the transient uncoupling scheme although both of them are inherently similar occasional coupling schemes.*

## VI. CAVEATS IN USING EIGENSPECTRUM

Although, in this paper, we have chosen to work with the (possibly complex) eigenvalues of $\tilde{\mathsf{J}}$, we could have used other quantities like local Lyapunov exponent[29,30] or eigenvalues of the symmetrized Jacobian[31]. However, for the purpose of our studies reported herein, all these three types of quantities are almost equivalent and so our choice is sufficient. In fact, our choice is also motivated and justified by the successful earlier usage[32] of the aforementioned complex eigenvalues for exploring synchronization related issues in the presence of large parameter mismatch between the unidirectionally and diffusively coupled chaotic oscillators. However, there are potential pitfalls in the aforementioned equivalence of the methods, that we discuss and address below.

When evolved for a small time $\Delta t$, the formal solution to Eq. (3) may be written as $\mathbf{x}_\perp(\Delta t) = [\mathsf{I} + \tilde{\mathsf{J}}(\mathbf{x_s}, \alpha; \mathbb{R}^3)\Delta t]\mathbf{x}_\perp(0) = \mathsf{L}(\Delta t)\mathbf{x}_\perp(0)$. It can be easily shown[14] that, to the first order in $\Delta t$, the singular values of $\mathsf{L}$ correspond to those non-contracting directions where the eigenvalues ($\gamma_i$'s, say) of the symmetrized Jacobian, $[\tilde{\mathsf{J}}(\mathbf{x_s}, \alpha; \mathbb{R}^3)^{tr} + \tilde{\mathsf{J}}(\mathbf{x_s}, \alpha; \mathbb{R}^3)]/2$, are nonnegative. The singular values of $\mathsf{L}$ can be seen as the local Lyapunov exponents[33]. Additionally, one may note that in the linearized dynamics of the transverse variables about the origin is strictly speaking a non-autonomous system since the Jacobian contains time dependent functions, $\mathbf{x_s}(t)$, that are, technically speaking, external 'parameters' in the system of equations for $\mathbf{x}_\perp$. Now problem is that the linear stability analysis (i.e., eigenvalues of the Jacobian) in non-autonomous systems may not give the correct local picture: Consider the example of the linear non-autonomous system,

$$\frac{d\mathbf{x}(t)}{dt} = \begin{bmatrix} -1 - 2\cos(4t) & 2 + 2\sin(4t) \\ -2 - 2\sin(4t) & -1 + 2\cos(4t) \end{bmatrix} \mathbf{x}(t), \quad (12)$$

where, quite deceptively, the two (degenerate) eigenvalues of the Jacobian are $-1$ and $-1$ for any $t$, although an exact non-trivial solution, $\mathbf{x}(t) = \exp(t)[\sin(2t), \cos(2t)]$, is divergent. However, the eigenvalues of the symmetrized Jacobian are $+1$ and $-3$ highlighting the divergent local dynamics. Thus, this is an additional benefit of working with the symmetrized Jacobian.

With this in mind, let us adapt the modified dynamic coupling to incorporate the idea of using symmetrized Jacobian. To this end, we recall Eq. (7) and keep in mind that, in what follows, all the discussion is specifically for this system. Rather than using $\lambda_1$, $\lambda_2$, and $\lambda_3$, let's use $\gamma_1$, $\gamma_2$, and $\gamma_3$ (such that $\gamma_1 \geq \gamma_2 \geq \gamma_3$ without any loss of generality) as introduced in this section. We take the specific cases where coupling strength, $\alpha$, is either 0.1 or 5.0 for the sake of comparison with the results reported in Sec. III and Sec. IV. We numerically find $\gamma_3$ is always negative and $\gamma_1$ is always positive so that if the dynamic coupling were to be applied, the dynamic coupling ($\mathsf{C_{dcs}}$) should be active for 100% of the attractor. However, we can better this by adopting another modified dynamic coupling scheme (analogous to the one introduced in Sec. III): dynamic coupling is turned on when $|\gamma_1| > |\gamma_3|$ or when $\gamma_2 > 0$, which ensures that the dynamics in the contracting directions is allowed on dominate over that in the non-contracting directions. This modified version of the dynamic coupling—rather weaker form of the dynamic coupling—requires the dynamic coupling term to be active only approximately at 30% and 19% of total accessible phase points on the respective attractors for $\alpha = 0.1$ and $\alpha = 5.0$ respectively. In this context, recall that when the modified dynamic coupling is employed using the eigenspectrum of the bare Jacobian discussed in Sec. III, the coupling term is active over $\sim 17\%$ and $\sim 2\%$ phase points for $\alpha = 0.1$ and $\alpha = 5$ respectively. It is evident that although the percentages differ, both the modified dynamic coupling schemes are improvement over the dynamic coupling scheme. Thus, the successes of the corresponding modified dynamics coupling schemes imply that *working with the eigenspectrum of the symmetrized Jacobian and working with the eigenspectrum of the bare Jacobian are qualitatively equivalent for the problem in hand.*

It must be commented that the aforementioned equivalence is partly due to the diffusive coupling employed in this paper. Using contrived couplings, it has been established[34] that having exclusively [35] negative real parts of all the eigenvalues of the Jacobian is not sufficient for guaranteeing a burst-free stable synchronized state. A rather obvious sufficient condition for the synchronization dynamics to be stable is that all the eigenvalues of the symmetrized Jacobian are negative everywhere on the corresponding attractor. This is, however, a very strict condition and moreover not a necessary one: many coupled systems—e.g., two diffusively coupled Rössler systems—do not have such Jacobians, and still have stable synchronized state.

## VII. SUMMARY AND CONCLUSIONS

We emphasize that our aim in this paper is not to introduce the transient uncoupling scheme which has already been done[11,15] in the literature for synchronization between identical chaotic systems and extended[18] for noisy nonidentical chaotic systems. Here, through the idea of the modified dynamic coupling devised in this paper, we have systematically scrutinized the transient uncoupling scheme and compared it with the on-off coupling scheme. We have found that the reason behind the synchronization effected by the transient uncoupling for high values of coupling strength, at which the individual oscillators do not synchronize when continuously coupled, can be traced to the set of spectra of eigenvalues of the Jacobians found at each point of the response subsystem's trajectory. We have shown that, in order to explain the success of the transient uncoupling, the eigenvalues with the minimum real parts of the local Jacobians must be taken into account in addition to the eigenvalues with the maximum real parts. This decisive fact was not realized in earlier publications on the topic of the transient uncoupling induced synchronization. In this context, we have introduced the concept of the transient decay and illustrated how it plays a crucial role behind the effectiveness of the transient uncoupling. Additionally, we have employed the transient uncoupling successfully to synchronize to the noisy identical hyperchaotic systems and understood the phenomenon in terms of the eigenspectra of the local Jacobians. We should stress here that, as far as the numerical experiments for the transient uncoupling presented in this paper are concerned, we have first arbitrarily fixed the coupling region, $\mathbb{A}$, and then turned the coupling on or off depending on whether the trajectory is inside $\mathbb{A}$ or outside. What we have explained in this paper is, why certain $\mathbb{A}$ works and certain ones don't; also, had one desired to pick an optimal $\mathbb{A}$ without doing trial-and-error, then what one should do.

We also reiterate that we have constructed a new synchronization scheme that is a significant improvement over the standard dynamic coupling scheme. Although mostly used as a tool to understand the efficiency of the transient uncoupling, this new method—termed modified dynamic coupling scheme—in itself is an interesting result. This scheme is applicable to the chaotic and the hyperchaotic systems alike, and is robust against noise as well. As an illustration of the surprising effectiveness of the modified dynamic coupling over the dynamic coupling, we have seen that when the dynamic coupling effects synchronization, the dynamic coupling is active at 14391 points out of the total 15499 phase points (approximately 93%) on the attractor at $\alpha = 0.1$, whereas at $\alpha = 5.0$ the coupling must always active. In contrast, for the same sampling rate, with the modified dynamic coupling employed, the coupling can be kept active only at 2614 and 317 points (i.e., approximately 17% and 2% respectively) when $\alpha = 0.1$ and $\alpha = 5.0$ respectively for bringing synchrony between the two chaotic oscillators. This highlights remarkable reduction (of approximately 76% for $\alpha = 0.1$ and 98% for $\alpha = 5.0$) in the number of phase points at which the dynamic coupling should be on in order to successfully bringing about chaotic synchronization.

A possible criticism of the work presented in this paper could be that while defining the two parameters, viz., $R_1$ and $R_2$, we have only considered the number of phase points and not weighted them with the magnitudes of the real parts of the corresponding eigenvalues. While this is not an unfair point, it doesn't add any new valuable practical insight. In fact, we have found that our conclusions do not change much qualitatively, partly because we are not giving any quantitative threshold of synchronization for these two parameters. Any attempt to find such (universal) thresholds is futile as they are bound to change with any change in the type of oscillators being coupled. This, however, does not take the sheen away from $R_1$ and $R_2$ as their usages highlight and exemplify how the local information about the eigenvalues can explain why the transient uncoupling works when it does.

We end by commenting that it is, in principle, straightforward to extend the ideas developed herein to the cases of the transient uncoupling employed on the diffusively coupled non-identical subsystems, the bidirectionally coupled subsystems, and the networks of oscillators.

## ACKNOWLEDGEMENTS

The authors thank Pushkar Khandare, Manu Mannattil, Malte Schröder, Aditya Tandon, and Marc Timme for fruitful discussions. The authors are grateful to Manohar K. Sharma for the help in making the plots presentable. The authors also thank the anonymous referees for their insightful comments. S.C. gratefully acknowledges financial support from the INSPIRE faculty fellowship (DST/INSPIRE/04/2013/000365) awarded by the INSA, India and DST, India.